# Ontology-based Solution for Building an Intelligent Searching System on Traffic Law Documents

Vuong T. Pham[1,2,4][a], Hien D. Nguyen[3,4][b,*], Thinh Le[3,4], Binh Nguyen[2,4][c], Hung Q. Ngo[5][d]
[1]*Faculty of Information Technology, Sai Gon University, Ho Chi Minh City, Vietnam*
[2]*Faculty of Mathematics and Computer Science, University of Science, Ho Chi Minh City, Vietnam*
[3]*University of Information Technology, Ho Chi Minh City, Vietnam*
[4]*Vietnam National University, Ho Chi Minh City, Vietnam*
[4]*Technological University Dublin, Dublin, Ireland*
*vuong.pham@sgu.edu.vn, hiennd@uit.edu.vn, 19520285@gm.uit.edu.vn, ngtbinh@hcmus.edu.vn, hung.ngo@tudublin.ie*

Keywords: knowledge base, searching system, traffic law, law on road traffic, legal document

Abstract: In this paper, an ontology-based approach is used to organize the knowledge base of legal documents in road traffic law. This knowledge model is built by the improvement of ontology Rela-model. In addition, several searching problems on traffic law are proposed and solved based on the legal knowledge base. The intelligent search system on Vietnam road traffic law is constructed by applying the method. The searching system can help users to find concepts and definitions in road traffic law. Moreover, it can also determine penalties and fines for violations in the traffic. The experiment results show that the system is efficient for users' typical searching and is emerging for usage in the real-world.

## 1. INTRODUCTION

Nowadays, transportation is a need for everyone. Almost every adult has a vehicle - the traffic is increasingly complicated, especially road traffic. In Vietnam, there are more than three million traffic law violations, with more than 14,500 traffic accidents in 2020 (National Traffic Safety Committee, 2020). Some cases have resulted in injuries or deaths. The reason for those cases is that people have low awareness of the rules of traffic law.

Ontology is an effective approach to representing knowledge (Jakus et al., 2013). This model has been used to organize knowledge in education and healthcare (Do et al., 2018). Moreover, several studies adopt ontologies to represent the knowledge of legal documents, while other studies use ontology to organize legal knowledge (Valente and Breuker, 1992, Fawei et al., 2019). However, they did not mention the traffic law for searching its content and determining penalties for violations.

This paper proposes a method for building the knowledge base for Vietnam road traffic law (Vietnam National Assembly, 2008, Vietnam Government 2019). This method is applied to construct a search system in this law. The designed system supports users in finding the content of the law related to their queries, and it can determine penalties for violations in road traffic via this law. In addition, the system helps to raise people's awareness about traffic law.

---

* Corresponding author
[a]  https://orcid.org/0000-0002-3879-9677
[b]  https://orcid.org/0000-0002-8527-0602
[c]  https://orcid.org/0000-0001-5249-9702
[d]  https://orcid.org/0000-0001-8246-8392
[d]  https://orcid.org/0000-0001-5249-9702
[d]  https://orcid.org/0000-0001-8246-8392

The primary value of the designed system is the ability to search for penalties and fines for road traffic offenses based on the keywords of the inputted query. Therefore, the system's knowledge base is organized as a relational ontology, which includes concepts, entities, their relations, and the rules of Vietnam Law on road traffic. In order to do that, the knowledge domain about road traffic law is collected and classified into knowledge components: concepts, relations, and rules.

The following section presents related works for constructing relational ontology, especially in the law domain, and several search systems on legal documents. Section 3 proposed an improved model of Rela-model to represent the knowledge of the road traffic code in Vietnam. Section 4 builds an architecture and searching problems of an intelligent querying system on Vietnam traffic code. The designed system can support finding the content of law related to the query and penalties for road traffic offenses. The last section concludes the results of this paper and gives some future works.

## 2. RELATED WORK

There are many studies to organize legal knowledge. For example, Valente and Breuker (1992) stated three approaches for the legal knowledge base: the logic approach, the case-based approach, and the pragmatic approach. Those approaches are used to build legal ontologies and documents for data-retrieving systems (Sator et al., 2011).

Ontology LIDO for Legal Informatics Document is built based on the standard CEN Metalex (Sartor et al., 2019). It represents legal actions that affect the document, the legal temporal events, the structure of the legal resource, and the semantic structure of organization of legal documentss.

Ngo et al. (2021) proposed a method of data augmentation based on legal domain knowledge for the legal textual entailment. This method is used to design a system for Vietnamese legal text processing. Nguyen et al. (2022c) also proposed a training data augmentation procedure and an unsupervised embedding learning method to retrieve the legal document. However, those proposed methods only show the articles of a specified query and does not use legal knowledge to explain its results clearly.

Pham et al. (2019) built an ontology-L for representing the Law of Public Investment and designed a consultant system for estimating the costs of a project based on this law. In addition, an intelligent chatbot was designed to tutor some administrative procedures in printing licensing based on the ontology Rela-Ops model (Nguyen et al., 2020a). However, those methods are challenging to apply in searching the content of a law document related to the working domain.

There are some legal search systems in Vietnam, such as the National Database of Legal Documents (2022) of the Ministry of Justice and law library (2022). However, these systems generally only allow users to search for documents or entities with keywords. However, they cannot help users find a deeper search for legal documents in the real world. For example, in traffic law, users need to search for penalties and fines for a violation based on rules in the legal document. Therefore, the current systems are not suitable for supporting users in practice.

This study tends to build an intelligent search system based on the ontology of the Vietnam road traffic code. This ontology can be used to represent the content of the law code and to deduce based on the inference rules extracted from the code.

## 3. KNOWLEDGE BASE OF VIETNAMESE TRAFFIC LAW

### 3.1 The structure of the Vietnamese law on the road traffic

This section gives more details about the structure of Vietnamese law on road traffic and the knowledge model of the system. Through Vietnam National Assembly (2015), the system of legal documents in Vietnam has the following levels:
1. The highest validity is Constitution;
2. Codes/Laws and resolutions of National Assembly;
3. Sub-law documents for instructing the detail of the law established by National Assembly.

In general, a law document has a structure with three parts: heading, content, and ending. The heading shows the national name, the crest, number, and sign of the document, enact place and date, type and name of the document, and the basis of the document. The content is a list of parts, chapters, articles, clauses, and points. The ending is the signing of the person that implements the document.

Inside the content, part is the highest level, then, in order are chapters, sections, articles, clauses and points. Through (Vietnam Ministry of Justice, 2011), based on the type of document, there will be different structures, for example some documents have chapters, articles, clauses, and points but there is no section. Each part, section, or chapter defines a different factor. Below chapter are articles and clauses which are used to define concepts, principles, penalties, or regulations. If a clause needs more than a sentence to define it, there will be several points in addition to it.

Concepts in legal documents have two parts, concept names and their definitions. For the offences, each principle, penalty, or regulation which are defined in articles and clauses of the legal document, they always have the subject (the person or organization that participate or engage in the event), a fact (or action) and penalties if there is any.

In particular, the Vietnamese traffic law has the same structure as stated. Two legal documents currently implement and have most effect in the social are: Law on road traffic (Vietnam National Assembly, 2008) which prescribes interpretation of concepts, road traffic rules, regulations for vehicles and users on the road traffic; Decree of Administrative of penalties for road traffic offences and rail transport offences (Vietnam Government, 2019) (known as Decree 100) which states penalties and fines for administrative violations of road traffic. In addition, there is National Technical Regulation on Traffic Signs and Signals (Vietnam Ministry of Transport, 2019) to define and describe the road traffic signs.

## 3.2 Knowledge model for road traffic law

Ontology Rela-model is a useful ontology representing the knowledge of relations. This model includes three components about concepts, relations between concepts (Nguyen et al., 2015). It is effective to represent knowledge domains in education, consultant the finance method based on the investment law. Rela-model includes three components which are used to represent concepts, relations between concepts and inference rules of the knowledge domain.

For representing the knowledge of a legal document, Rela-model has been improved the structure of its concept-component being suitable the legal domain (Nguyen et al., 2022a). The knowledge model for Vietnamese road traffic law is based on the concepts or entities and their relations. Each relation of them defines an action or event of road traffic. Based on those relations and rules of law on road traffic, the issues about retrieving the information of offences and their penalties have been also proposed.

**Definition 2.1:** The knowledge model for representing the legal domain of road traffic is improved from ontology Rela-model, named *Traffic-Law model*. This model consists of three components as follows:

$$(C, R, Rules)$$

In which, **C** is the set of concepts or entities of road traffic law, **R** is the set of relations between concepts/facts, **Rules** represent the inference rules to specify the relation between concepts or determine offences and their penalties. The structure of Traffic-Law model is summarized as Figure 1.

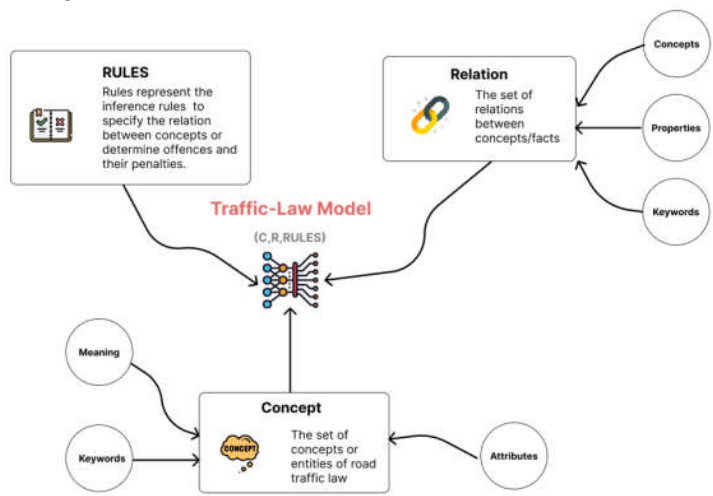

Figure 1. The Traffic-Law model.

Set **C** is the set of concepts and entities in road traffic law. There are three kinds of concepts in C: users and vehicles of road traffic; traffic signs and signals; road infrastructure. Based on those kinds, each concept $c \in$ **C** has the structure:

(*Name, Meaning, Attrs, Keywords*)

where, each element has the type and meaning for specifying the corresponding concept as Table 1:

Table 1. Structure of a concept

| Element | Type | Meaning |
|---|---|---|
| *Name* | Text | Name of the concepts |
| *Meaning* | Text | Meaning of the concepts. |
| *Attrs* | Dict | List of attributes of the concepts. |
| *Keywords* | Set | Set of keywords determined or related to the concepts. |

Example 1: The concepts *"Electric motorcycle"* in (Vietnam Government, 2019) is described:

| Element | Content |
|---|---|
| *Name* | Electric motorcycle |
| *Meaning* | "a two-wheel vehicle operated by an electric engine with power not exceeding 4 kW and maximum speed not exceeding 50 km/h" |
| *Attrs* | *Attrs* = [*kind, type, legal*]<br>• *kind*: road traffic vehicle<br>• *type*: two-wheel vehicle<br>• *legal*: [Article 3, Clause 1, Point d, Decree No. 100/2019/ND-CP] |
| *Keywords* | Motorcycle; electric; two-wheel vehicle |

Set **R** is the set of relations between concepts in set **C**. These relations determine a specific fact or an action of the road traffic code. Each relation $r \in$ **R** has the structure:

(*Name, Conc, Meaning, Prop, Keywords*)

where, each element has the type and meaning for specifying the corresponding relation as Table 2:

Table 2. Structure of a relation

| Element | Type | Meaning |
|---|---|---|
| *Name* | Text | Name of the relation. |
| *Conc* | List | List of parameters as concepts of the relation. |
| *Meaning* | Text | Meaning of the relation. |
| *Prop* | Set | Set of properties of the relation. This study only mentions two main properties on a binary relation: transitive and symmetric. |
| *Keywords* | Set | Keywords of the relation. |

Example 2: The relation "*comply*" of two concepts "*car*" (or "*car-like vehicles*") and "*traffic light*", denoted *comply* (*car*, *traffic light*), means "Operators of car and car-like vehicles failed to comply with the traffic lights". Its keywords are "*comply*", "*over*".

Set **Rules** is a set of inference rules. Those rules deduce relations between concepts or determine offences based on road traffic law. Each rule $r \in$ **Rules** has the form

$$u(r) \rightarrow v(r)$$

where, $u(r)$ is the hypothesis facts of rule $r$ and $v(r)$ is the result of rule $r$.

The **Rules**-set is classified two kinds of rules:

**Rules** = **Rule**$_{infer}$ ∪ **Rule**$_{offence}$

In which, **Rule**$_{infer}$ is the set of rules inferring the relation between concepts, and **Rule**$_{offence}$ is the set of rules determining offences and penalties.

## 3.3 Some problems for searching on traffic law

Using the improved Rela-model, the knowledge base for road traffic law has been organized. Based on this knowledge base, the problems for searching on the law document are studied. There are two issues for searching on law, which are searching for the concepts or definition of the law, especially the law explanation, and determining offences and their penalties and fines through the law document. To do this, two searching problems need to be solved for designing the intelligent searching system on the law document:

**Definition 2:** The searching problems of an intelligent searching system Traffic-Law model are:
- *Problem 1:* Extracting the keywords from the inputted query to search the concepts and relations in the legal knowledge base related to the keywords.
- *Problem 2:* Retrieve the knowledge from the knowledge base matching extracted concepts and relations.

For solving Problem 1, the inputted query needs to be classified. The input can be classified into two kinds: query about meaning of a concept ("*what is*?") and query about the penalties & fines of an offence ("*how much*", "*penalty*", "*fines*"). After that, from the kind of the query, its main keywords are extracted. In addition, some similar words for extracted keywords are also achieved. The similar keywords can be collected from legal document sources, experts (as lawyers or legal lecturers), or from dictionaries. With extracted keywords and determined similar words, concepts related to those keywords are determined by using rules in **Rule**$_{infer}$. The process also finds inference rules used to deduce concepts and their relations.

**Algorithm 3.1:** Given a law document d which is already represented using ontology based on Traffic-Law model.

**Input:** The knowledge base $\mathcal{K}$ = (**C, R, Rules**) as Traffic-Law model.
Query *q*.

**Output:** A set of keywords, relations, and rules retrieved from query *q* and knowledge $\mathcal{K}$

Algorithm:
  **Step 1:** Classify the query using Vietnamese NLP toolkit
  **Step 2:** Extract keywords from the query *q* and find similarly words based on the knowledge base $\mathcal{K}$.
  $$W := keywords(q)$$
  **Step 3:** Classify the kind of query based keywords in *W*.
  **Step 4:** Expands W with similar keywords collected from legal sources.
  **Step 5:**
  $G$ := {} // Set of concepts
  $P$ := {} // Set of rules
  For each keyword $w \in W$ do
  • Using **Rule**$_{infer}$ to search concepts and rules related to w.
  • From found concepts, determine required keywords and add them to *G*.
  • Add rules to *P* if not exists
  **Step 6:** Return (*G*, *P*) are results of found keywords and rules.

For solving Problem 2, after identifying the concepts and relations, the article of legal documents that states the offence is found by using rules in **Rule**$_{offence}$. Then, the information, penalties, and fines of it are retrieved through the specified content of law in the knowledge base. The process for solving this problem is as follows:

Given the knowledge base $\mathcal{K}$ of road traffic law in legal documents as Traffic-Law model. This algorithm will determine the information, penalties, or fines of an inputted query *q*.

**Algorithm 3.2:** Given a law document *d* which is already represented using ontology based on Traffic-Law model

**Input:** The knowledge base $\mathcal{K}$ = (**C, R, Rules**) as Traffic-Law model, and a query *q*.
**Output:** Information, penalties, and fines of road traffic offence for query *q*.

Algorithm:
  **Step 1:** Retrieve set of keywords *G* from query *q* based on Algorithm 3.1
  *Concept* := {$c \in$ **C** | *c* related to keyword in *G*}
  **Step 2:**
  *Knowledge* := {}
  For each concept $c \in$ *Concept* do:
  • Using rules in **Rule**$_{offence}$ to find the offence in the knowledge base $\mathcal{K}$.
  • Retrieve the information, penalties, and fines of the determined offence from the specified law document.
  • Update the results into *Knowledge*.
  **Step 3:** Return *Knowledge*.

## 4. THE SEARCHING SYSTEM OF VIETNAMESE LAW ON ROAD TRAFFIC

### 4.1 Requirements of a searching system on legal documents

The intelligent searching system on legal documents needs to be supported the understanding of users about the legal domain. In road traffic law, moreover, the ability for solving of necessary issues of the searching system, this system has some criteria of intelligent software evaluation in searching (Nguyen et al., 2020b, Giakoumakis and Xylomenos, 1996):

  o *Portability:* This is the level of difficulty to work with the same project with different machines.
  o *Installation:* The requirements of software, hardware for the simulator, and how straightforward is the installation in a supported system.
  o *Usability:* this criterion shows whether the content is suitable and detailed with the current law domain and whether it is updated and easily to use in the practice.
  o *Understandability:* this is one of the most important characteristics of intelligent law searching software quality. This system has to help users understand the law content in legal documents. It can influence users' feelings about software and reliability of software evolution in reuse or maintenance.

Besides, the process for building this system is worked through the constructing of a knowledge-based system (Nguyen et al., 2022b). At first, the databse of traffic regulations will be collected, and orgnaized by Traffic-Law model as the knowledge base of this system. After that, the searching mechanism is designed through problems on traffic law searching and their alogrithms. Finally, the user interface and testing of this system will be processed.

## 4.2 The dataset of traffic regulations

The traffic regulation dataset is a combination of 2 documents:
1. Vietnam National Assembly, Law on Road Traffic (known as 23/2008/QH12).
2. The Decree of Administrative of penalties for road traffic offences and rail transport offences (Vietnam Government, 2019), abbreviated as Decree 100.

From both documents, there are 175 articles collected. The general structure of these documents is: Chapter – Section – Article – Clause – Point. Traffic-Law model is used as an ontology to represent this knowledge.

By default, questions about Vietnamese transportation are classified into many intents. There intents include but not limited to:

• Querying about concepts: These queries ask definitions of concepts in the law. The system extracts the apporiate article for the required concept.

• Querying about penalties: These queries ask about the penalty or fines for a traffic violation, such as running the red light, driving contrariwise, etc.

• Querying about procedures: The system give a proceduce in traffic law, such as fine payment procedure, the procedure for issuing driving licenses, etc.

• Querying about signs: This function support user to retrieve the information of an inputted sign. This function related to image processing.

However, because the scope of this study, only the kinds of querying about concepts and penalties are focused in this paper. There are 160 practical collected queries related to road traffic regulation. These queries will be augmented and used for training query intent classification in Problem 1.

Table 3: Query Classification

| Class | Meaning | Quantity |
|---|---|---|
| Concept | Require identifying the meaning of a concept. | 54 |
| Penalties | Require identifying the fine of an offence. | 83 |
| Out of scope | Queries that do not belong to above kinds. | 23 |
| Total | | 160 |

## 4.3 The architecture of the searching system on the traffic law

The architecture of the search system on traffic law is presented in Figure 2. The system consists of the user interface, the knowledge base, and the search engine.

The knowledge of the road traffic codes is collected from (Vietnam Government, 2019, Vietnam National Assembly, 2008). These facts and entities of those documents are organized as a knowledge base by the improved ontology Rela-model and stored inside a graph database. The similar words are manually established via the collection of intellectual experts and their experiences.

When a user inputs the query, the search engine will execute the extract keywords tasks by Problem 1, which are classifying the query, checking typo, removing stop words, checking synonyms, and checking equivalent keywords, to generate the query values. From the extracted keywords, the similar words will be determined through the knowledge base Traffic-Law model. Those are used to search the necessary knowledge by using inference rules of the knowledge. In addition, their penalties and fines are also retrieved by Problem 2. The result will be ranked by the search engine before showing it in the user interface.

## 4.4 Testing Results

Based on the knowledge base that has been organized in Section 3 and the proposed architecture in Section 4.2, an intelligent searching system on Vietnam Road traffic law is designed. This section presents some testing results of the system through some kinds of inputted queries.

Example 3: The inputted query $q_1$ = "*What is motorcycle*?"

The system will extract keywords from the query $q_1$: "*What is*", "*motorcycle*". From that, it returns the results as follows:

"*Motorcycle means a motor vehicle that has two or three wheels with a cylinder capacity of 50 cm$^3$ or higher, maximum speed over 50 km/h, and net weight not exceeding 400 kg.*"

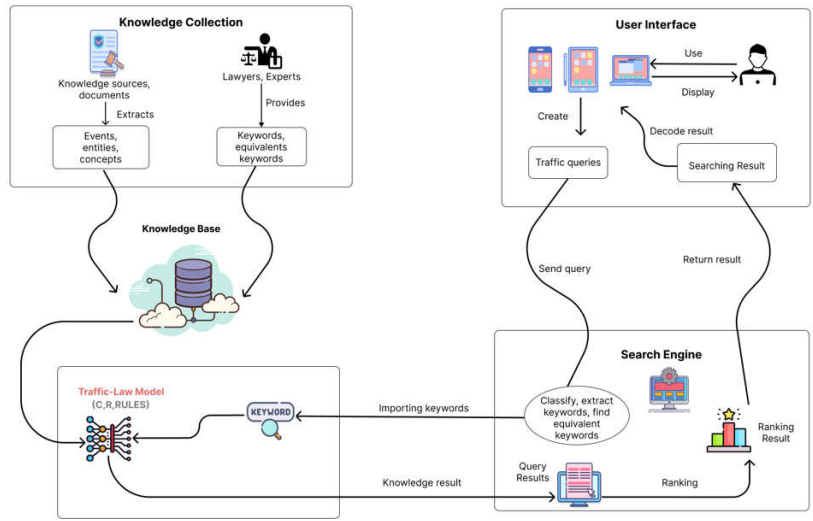

Figure 2. The architecture of an intelligent searching system on the Vietnam road traffic.

The word "*what is*" is used to classify the query into the kind of declaring the meaning of a concept. The keyword "*motorcycle*" helps to find the concept. The result is retrieved from Article 3, Clause 3.31 of National Technical Regulation on Traffic Signs and Signals (Vietnam Ministry of Transport, 2019).

Example 4: The inputted query $q_2$ = "*The fines of operator of motorbike driver who does not wear helmet*"

The keywords of the query $q_2$ are *"fines", "not wear", helmet", "operator of motorbike"*. The word "*fines*" is used to classify the query into stating penalties and fines of offences. The word "*operator of motorbike*" consists of "*motorbike*" that is similar to the word "*motorcycle*". The word "*helmet*" is in the keywords of the concept "motorcycle helmet". Hence, the concepts of the query $q_2$ are "*operator of motorcycle*" and "*motorcycle helmet*". The relational keyword is "*not wear*". With the concepts and relation, the rules were used to match them and find the result.

The result is returned:

"*Through article 6, Decree 100/2019/ND-CP: Penalties imposed upon operators of mopeds and motorcycles (including electric motorcycles) and the like violating road traffic rules.*

*2. A fine ranging from VND 200,000 to VND 300,000 shall be imposed upon a vehicle operator who commits any of the following violations:*

*i) The operator or the passenger on the vehicle does not wear a motorcycle helmet or does not wear it properly;*"

The designed system can do some common searching on road traffic law. It is effective in finding usual penalties and fines from road traffic law. This system was tested on a set of 95 queries about the road traffic codes. The results are shown in Table 4:

Table 4. Results for testing of queries

| Kind | Quantity | Correct | Rate |
|---|---|---|---|
| Queries about concepts / definitions | 54 | 42 | 78% |
| Queries about penalties and fines | 83 | 61 | 73% |
| **Total** | **137** | **103** | **75%** |

## 5. CONCLUSION AND FUTURE WORK

This paper proposed an ontology-based model for representing legal knowledge in the Vietnam road traffic codes. This model is improved based on ontology Rela-model in the structure of concepts, relations, and inference rules. Through the designed knowledge base, several searching issues on the Vietnam road traffic codes are proposed, such as extracting keywords and inferring the matched result for inputted query. Moreover, the architecture of an intelligent search system on road traffic law has been constructed. This system can do several common search queries, such as finding concepts/definitions in the law and determining penalties for violations in the road traffic. At the moment, most knowledge is collected by manual collection method. The next work is the improvement of the collection method within by using an automatic method.

In the future, the system can be involved other legal aspects such as commercial law, civil law, etc. Further, the system can be used to provide an e-learning system for legal aspects. The abilities to use AI to identify entities and concepts from an image or use voice recognition to identify searching input are also features considered to add more to the system.

## ACKNOWLEDGMENT

This research was supported by The VNUHCM-University of Information Technology's Scientific Research Support Fund.